\documentclass[useAMS,usenatbib,usegraphicx,letters]{mn2e}

\usepackage{psfig}
\makeatletter

\newcommand{\Rmnum}[1]{\expandafter\@slowromancap\romannumeral #1@}
\makeatother

\title[local photoionizing flux and gas cooling]{Stars quenching stars: how photoionization by local sources regulates gas cooling and galaxy formation}
\author[S. Cantalupo]{Sebastiano Cantalupo\thanks{E-mail:
cantal@ast.cam.ac.uk}\\
Kavli Institute for Cosmology, Cambridge and Institute of Astronomy, Madingley Road, Cambridge CB3 0HA, UK}
\begin{document}

\date{Accepted 2009 December 20. Received 2009 December 10; in original form 2009 September 24}

\pagerange{\pageref{firstpage}--\pageref{lastpage}} \pubyear{}

\maketitle

\label{firstpage}

\begin{abstract}
Current models of galaxy formation lack an efficient and physically constrained 
mechanism to regulate star formation (SF) in low and intermediate mass galaxies. 
We argue that the missing ingredient could be the effect of photoionization by local sources 
on the gas cooling. We show that the soft X-ray and EUV flux
generated by SF is able to efficiently remove the main coolants (e.g., He$^{+}$, O$^{4+}$ and Fe$^{8+}$)
from the halo gas via direct photoionization. As a consequence, the cooling and accretion time of the
gas surrounding star-forming galaxies may increase by one or two orders of magnitude. For a given halo mass and
redshift, the effect is directly related to the value of the star formation rate (SFR). Our results
suggest the existence of a \emph{critical} SFR above which ``cold'' mode accretion
is stopped, even for haloes with $M_{vir}$ well below the critical shock-heating mass suggested
by previous studies. 
The evolution of the critical SFR with redshift, for a given halo mass, resembles the respective
steep evolution of the observed SFR for $z<1$. This suggests that
photoionization by local sources would be able to regulate gas accretion and star formation,
without the need for additional, strong feedback processes.
\end{abstract}


\section{Introduction}

Despite many successes, current models of galaxy formation have still several difficulties in reproducing
the galaxies as we see them today.  
The gas ``overcooling'' in cosmological simulations is considered as one of the main 
sources of this discrepancy with observations.  
Feedback mechanisms like supernovae (SN) or galactic winds heating have been advocated in order to 
suppress gas cooling and star formation (Dekel \& Silk 1986). However, 
numerical models including these feedback effects have not been able, so far, 
to obtain realistic galaxies and to reproduce the steep decline of the star formation rate 
at $z<1$ (see e.g., Crain et al. 2009).

While this may be related to numerical limitations (that force the introduction of \emph{ad hoc} prescriptions), 
it is worthwhile to re-consider on a physical basis
the gas cooling mechanism and how the galaxies get their gas. As originally pointed out
by Silk (1977) and Rees \& Ostriker (1977), galaxy-formation is regulated by the relative value of
the cooling and collapse timescales. Gas in haloes with masses below $10^{11.5}\mathrm{M_{\odot}}$,
cools so efficiently that it always accretes, roughly, at the free-fall rate. This ``cold'' mode
accretion (as opposite to the ``hot'' mode, where gas is stably shock-heated to the halo virial
temperature) has been confirmed recently by more sophisticated analytical and numerical models (Dekel \& Birnboim 2006, 
Ocvirk et al. 2008 and references therein).
The physics of gas cooling is determined by the abundance of the ions that act as coolants for the free
electrons in the gas. Thus, every process that changes these abundances will act directly on the galaxy-formation
timescales. This simple idea has been the base of the seminal work by Efstathiou (1992), where it is
clearly illustrated how the cosmic photoionizing background suppresses the cooling due to H$^0$ and He$^{+}$
for low density gas (see also Wiersma et al. 2009). 
This effect is important to suppress the formation of (dwarf) galaxies within dark-matter haloes 
less massive than $10^{9}\mathrm{M_{\odot}}$ but it does not substantially
affect cooling and accretion of virialized, dense gas in more massive haloes (indeed, the majority of 
numerical models already include this effect in the cooling function). 

In this Letter, we consider an ingredient neglected so far: the effect of the \emph{local} photoionizing flux 
from star formation on gas cooling and accretion. 
Star-forming galaxies are conspicuous sources of soft X-ray and EUV photons. In Section 2, we discuss their spectral energy distribution. 
In Sections 3 and 4, we calculate the effect of the local photoionization flux on gas cooling and galaxy formation. We conclude in Section 5.
We use a flat $\mathrm{\Lambda}$CDM cosmology with $h=0.7$, $\Omega_{m}=0.27$,
$\Omega_{\mathrm{\Lambda}}=0.73$, $\Omega_{b}=0.046$, $X_{\mathrm{He}}=0.25$, 
and we denote with ``pKpc'' distances in physical kpc.
 
\begin{figure*}
\centerline{
\includegraphics[width=52mm]{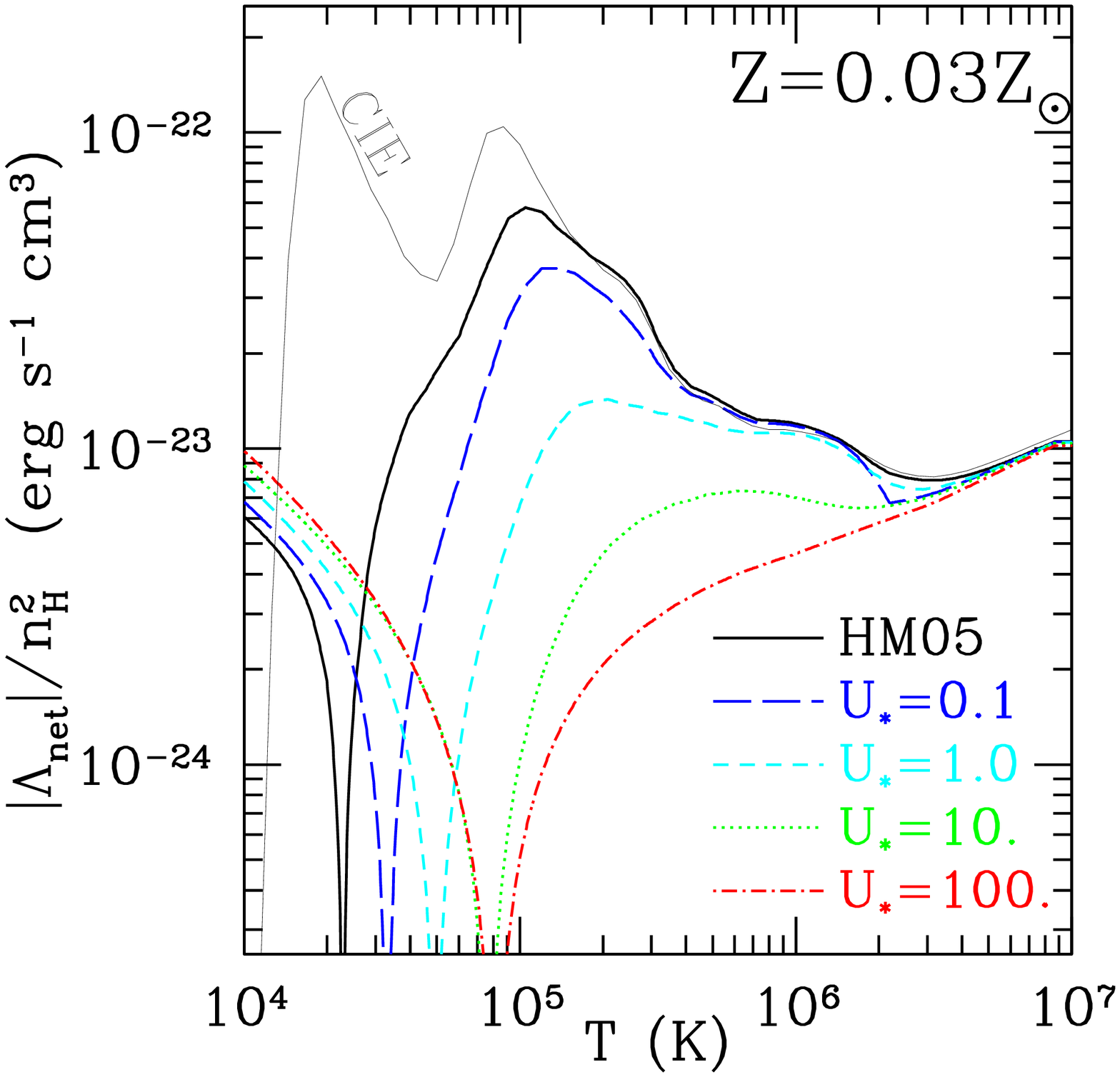}
\includegraphics[width=52mm]{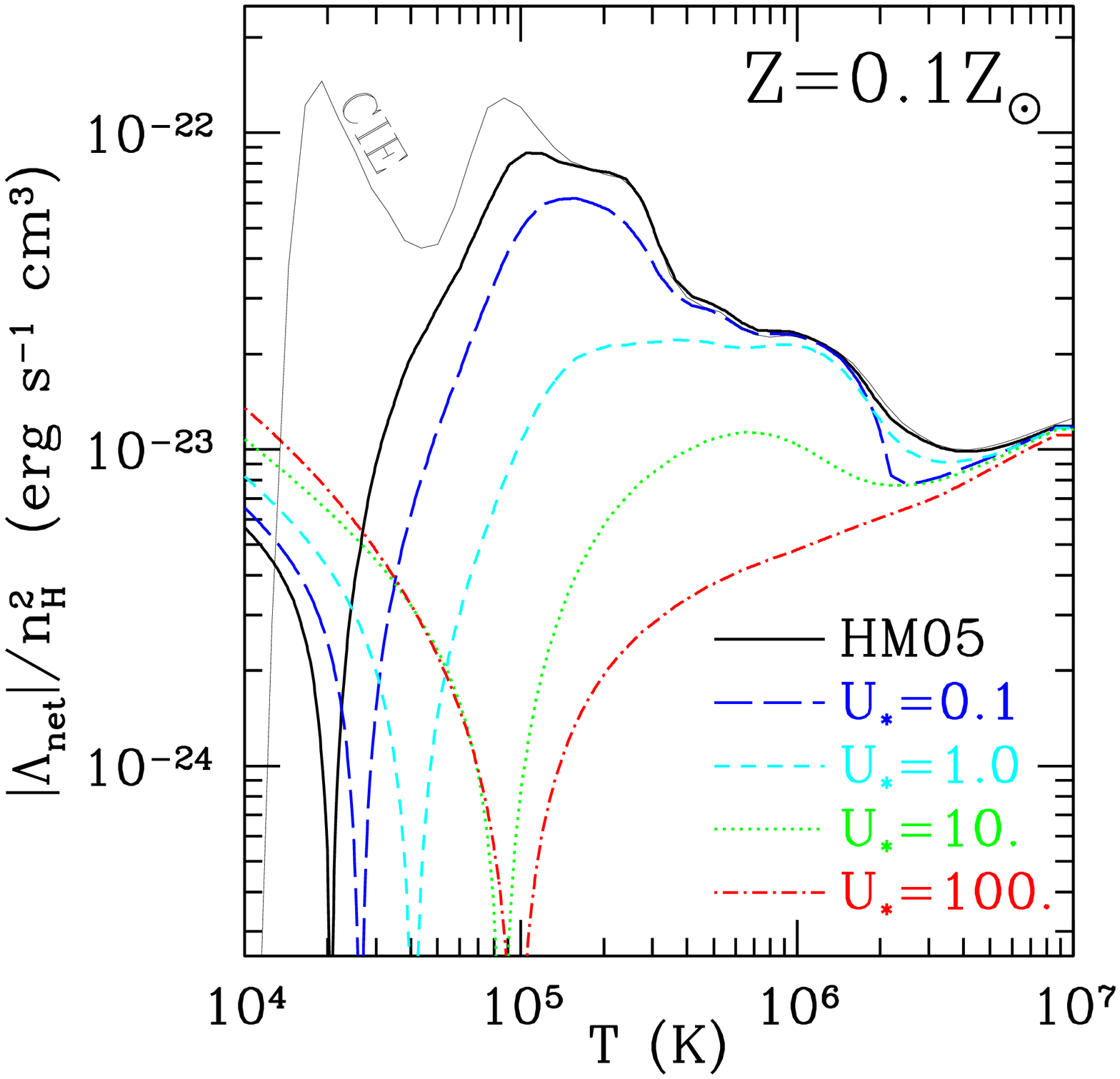}
\includegraphics[width=52mm]{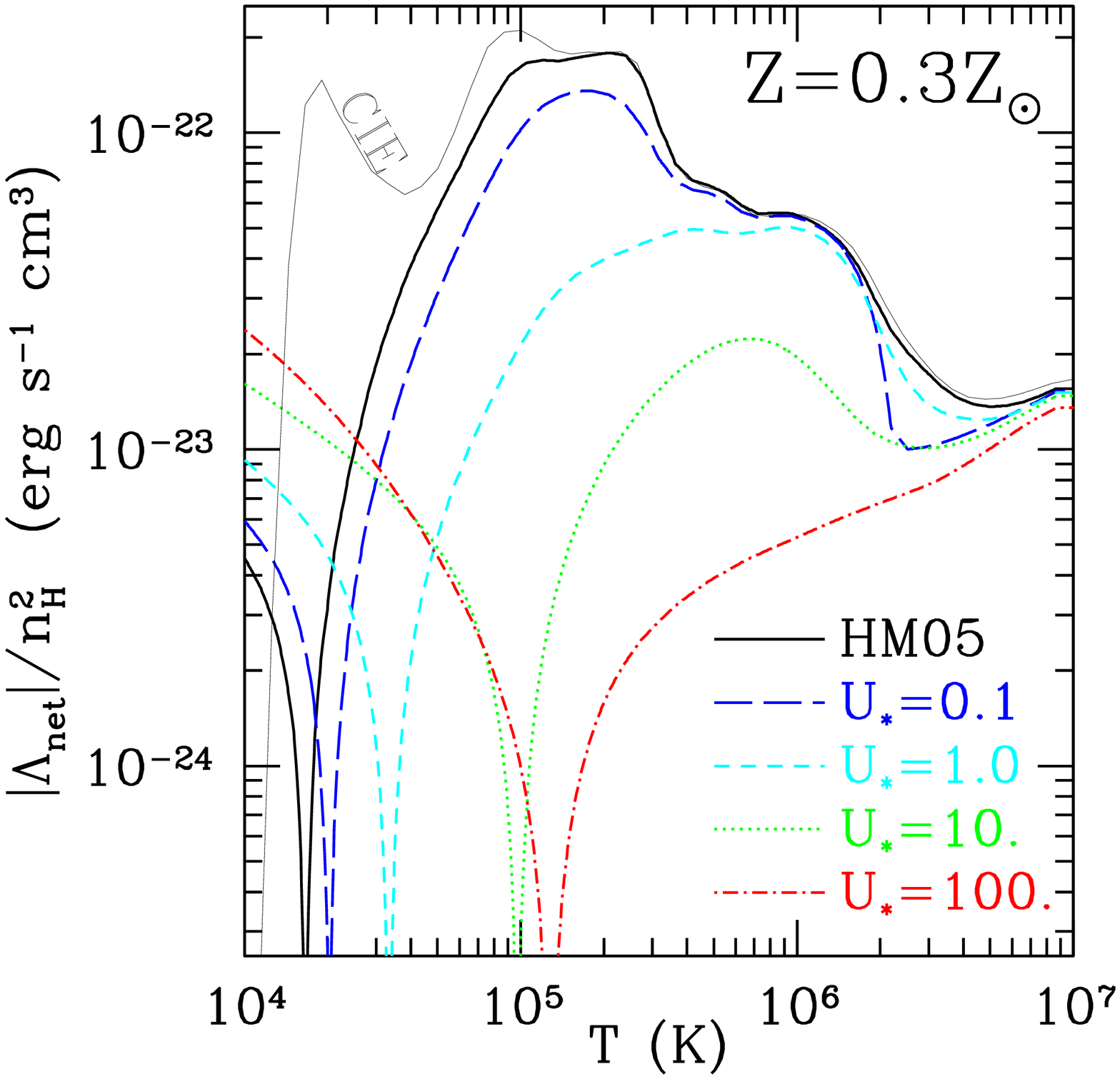}
}
\caption{Net cooling rates (defined as the absolute value of the radiative cooling minus the photo-heating rate) 
as a function of gas temperature for three different metallicities (see labels in panels). 
The gas density is $n_{H}=1.6\times10^{-3}$ cm$^{-3}$ ($\Delta\sim10^3$ at $z=1$). 
The light solid lines show the cooling rate in the Collisional Ionization Equilibrium (CIE) case, when
no photoionizing flux is present.
The dark solid lines represent the net cooling rate (note that heating dominates over cooling
at low temperatures) in the presence of the cosmic UV background (HM05) alone at $z=1$, 
while the other lines show the effect of adding the local
photoionizing flux from star formation with $\log(U_{*})=[-1,0,1,2]$ (see labels). 
These values can be directly translated to SFRs using Eq.\ref{Ustar} or Eq.\ref{Ustar2}.}
\end{figure*}

\section{The photoionizing flux of star-forming galaxies}\label{flux_sec}

 The onset of star formation episodes within galaxies produces a large amount of ionizing photons 
associated with a range of different processes. The EUV part of the spectrum
is dominated by the continuum of massive, hot and young stars. Conspicuous X-ray emission is produced by
the mechanical energy released into the Interstellar medium (ISM) during the starburst (Heckman et al. 1995, 
Strickland et al. 2004), by
non-thermal processes associated with supernovae explosions, and by massive X-ray binaries (Grimm et al. 2003, 
Persic et al. 2004). In particular, the soft X-ray flux correlates well enough with the star formation rate (SFR)
obtained from other indicators (e.g., the far infrared luminosity) to be used as a proxy to derive the
SFR for actively star-forming
galaxies (e.g., Rosa-Gonz\'alez et al. 2009 and references therein).

For this reason, we use the spectral energy distribution (SED) derived from models that are
calibrated empirically to reproduce 
the relation between SFR and soft X-ray emission. 
In particular, we use the model of Cervi{$\mathrm{\tilde n}$}o et al. 2002 (http://www.laeff.inta.es/users/mcs/SED), that
is consistent with observations assuming: i) relatively young star formation
episodes (age of about 5 Myrs) and, ii) a few per cent efficiency ($\sim5$\%) in converting mechanical energy
from the starburst into ISM heating and X-ray emission 
(Mas-Hesse et al. 2008).   
Assuming a Salpeter IMF and $Z=\mathrm{Z_{\odot}}$, the relation between the SFR and the soft X-ray (0.5-2.0KeV) emission 
for this model is: 
$L_{\mathrm{softX}}=3\times10^{40}$[SFR/(1 $\mathrm{M}_{\odot}$ yr$^{-1}$)] erg s$^{-1}$.
 We use this relation to calibrate the overall normalization of the intrinsic SED as a function of the SFR. 

For typical star-forming galaxies, 
the soft X-rays flux is not substantially affected by ISM absorption. 
Thus, our
results would be independent on absorption effects
if the main coolants are ions with ionization potential corresponding to the energy of soft X-rays photons. As we show in Section 3, this is
indeed the case for gas with $T\geq 10^{5}$K. For lower temperatures, H$^0$ absorption becomes more important and we must consider
how many photons with frequency around the Lyman Limit can actually escape the ISM. 
In particular, we use a frequency-dependent escape fraction ($f_{esc}^{\nu}$) of the form
$f_{esc}^{\nu}=[f_{esc}^{LL}+(1-f_{esc}^{LL})\mathrm{exp}(-\tau_{\nu})]$,
where $f_{esc}^{LL}$ is the absolute escape fraction at the Lyman Limit, 
$\tau_{\nu}=\sigma_{\nu}N(\mathrm{H}^0)$ is the neutral hydrogen optical depth and $\sigma_{\nu}$ the corresponding cross section. 
This is compatible with a leaky-absorbers model (see, e.g. Ferland \& Mushotzky 1982) for the ISM with a covering
factor of $(1-f_{esc}^{LL})$. We fix $f_{esc}^{LL}=0.05$, compatibly with observational estimates for local starburst galaxies
(e.g., Bergvall et al. 2006, but see Grimes et al. 2009) and high-redshift LBGs (Steidel et al. 2001, Shapley et al. 2006). 
The value of $N(\mathrm{H}^0)$ determines the hardening of the spectrum around the Lyman Limit. We anticipate that 
this parameter has a little effect on our results and we fix $N(\mathrm{H}^0)=10^{20}$ cm$^{-2}$. 
Note that He and dust 
may be also an important sources of opacity. 
We found that these effects become significant in reducing the EUV flux 
for gas column densities higher than $N(\mathrm{H})=10^{21}$ cm$^{-2}$, assuming Galactic ISM abundances.  

The relation between the SFR and the total luminosity of ionizing photons ($\nu>\nu_{\mathrm{LL}}$) for the assumed SED 
is given by:
$L_{ion}=3\times10^{41} [\mathrm{SFR}/{1 \mathrm{M_{\odot}} \mathrm{yr^{-1}}}]\ \mathrm{erg\ s^{-1}}$ .
We stress that, even if $L_{ion}$ depends on $f_{esc}^{LL}$, the overall effect of the
ionizing photons on gas with $T\geq 10^5$ K is almost independent on this parameter\begin{footnote}{For instance, 
we have verified that a much lower escape fraction, $f_{esc}^{LL}=0.005$,
would change the cooling rates at $T\geq 10^5$K only by few percents with respect to the values presented in Section 3.}\end{footnote}.
Nevertheless, we use $L_{ion}$ (in terms of the ionization parameter\begin{footnote}{defined as the ratio between
the density of photons with $\nu>\nu_{LL}$ and the hydrogen density at a given distance from the source.}\end{footnote}) 
as a convenient way to parametrise our results in the rest of this Letter.

\section{Local photoionizing flux and gas cooling}\label{main_sec}

 We calculate now the effect of the ionizing photons
on the cooling of a gas patch in the galaxy halo.
 For the assumed SED (Section 2), the ionization parameter is given by:
\begin{equation}\label{Ustar}
U_{*}=0.03\times\left[\frac{\mathrm{SFR}}{1 \mathrm{M}_{\odot} \mathrm{yr^{-1}}}\right]\left[\frac{7\ \mathrm{pKpc}}{r}\right]^{2}
      \left[\frac{10^3}{\Delta}\right]\left[\frac{2}{1+z}\right]^{3}\ ,
\end{equation}
where $r$ and $\Delta$ are, respectively, the distance from the galaxy and the gas overdensity.
For comparison, the ionization parameter given by the cosmic UV background (Haardt \& Madau 2005, in preparation; HM05) 
at $z=1$ is $U_{HM}=7.5\times10^{-3}$ for $\Delta=10^3$.
Note, however, that the photon spectral distribution 
of the HM05 to the SED of the star-forming galaxies are substantially different, especially around the
peak of quasar activity ($z\sim2$).
Using the halo virial relations (see e.g., Mo \& White 2002 and DB06 for an useful compilation) 
and an isothermal gas profile (i.e., $\Delta(r) \propto r^{-2}$),  Eq.\ref{Ustar} translates into:
\begin{equation}\label{Ustar2}
U_{*}=0.03\times\left[\frac{\mathrm{SFR}}{1 \mathrm{M}_{\odot} \mathrm{yr^{-1}}}\right](M_{11}^{vir})^{-2/3}
      \left[\frac{0.05}{f_b}\right]\left[\frac{2}{1+z}\right]\ ,
\end{equation}
where $M^{vir}_{11}$ is the virial mass in units of $10^{11}$M$_{\odot}$ and $f_b$ is the gas fraction inside the halo. 
Note that, for an isothermal profile, $U_{*}$ is independent on the distance from the galaxy.

 How does the inclusion of the local photoionizing flux change the gas cooling? 
For a gas with primordial composition, we could estimate the effect on the two main coolants,  
H$^0$ and He$^+$ (mainly via HI and HeII Ly$\alpha$ emission), directly from the value of the ionization parameter.
However, in a more realistic situation, traces of metals are present and dominate the cooling at $T_5\geq 1$ 
(where $T_5=T/[10^5K]$)
via collisional excitations of several ions. In order to get the detailed ionic balance in the presence of 
an EUV/X-ray radiation field and, ultimately, the heating and cooling rates,
we use the publicly available package Cloudy (version 08.00 of the code last described by
Ferland et al. 1998), which contains most of the atomic processes that are important for our temperature
range.

The results of the photoionization model are presented in Fig.1, where 
we show the absolute value of net cooling/heating rate, $\Lambda_{net}=\Lambda-\mathcal{H}$ ($\Lambda$ and
$\mathcal{H}$ are, respectively, the radiative cooling and the photo-heating rate in erg cm$^{-3}$ s$^{-1}$) as a function
of the gas temperature for three different metallicities. 
In this plot, the gas density is fixed at $n_{H}=1.6\times10^{-3}$ cm$^{-3}$, corresponding to
$\Delta=10^3$ at $z=1$. In each panel, 
the thick solid line represents the net cooling rate (note that heating dominates over cooling
at low temperatures) in the presence of the HM05 alone, while the other thick lines show the effect of adding the
local photoionizing flux from star formation with $\log(U_{*})=[-1,0,1,2]$ (see labels). 
These values correspond roughly to $SFR\sim[3,30,300,3000]$ for $r\sim7$ pKpc (Eq.\ref{Ustar}) or $M^{vir}_{11}\sim1$ (Eq.\ref{Ustar2}). 

In the presence of a photoionization flux, the main cooling peaks are due to He$^+$ (at $T_5\sim1$), 
O$^{4+}$ (at $T_5\sim2$, via O {\small{\Rmnum 5}} 630$\mathrm{\AA}$ line), Ne$^{5+}$ (at $T_5\sim5$,
via Ne {\small{\Rmnum 6}} 400$\mathrm{\AA}$ line) and Fe$^{8+}$ (at $T_5\sim10$, via Fe {\small{\Rmnum 9}} 169$\mathrm{\AA}$ line).
Clearly, the local photoionizing flux strongly reduces the net cooling rate at $T_5\sim1$ for $U_{*}\geq1$ and at $T_5\sim10$
for $U_{*}\geq10$. This is due both to the suppression of coolant ions via photoionization and the injection of 
energy following the photoionization process (i.e., photoheating). In the extreme case of a highly overionized medium (i.e., 
$U_{*}=100$, dot-dashed lines in Fig.1) all the coolant ions are removed and the main cooling process 
if free-free emission, which is highly inefficient at low temperatures, but is important for $T\geq10^7 K$ gas
(e.g., the Intra Cluster Medium).

\begin{figure}
\centerline{\vbox{
  \psfig{file=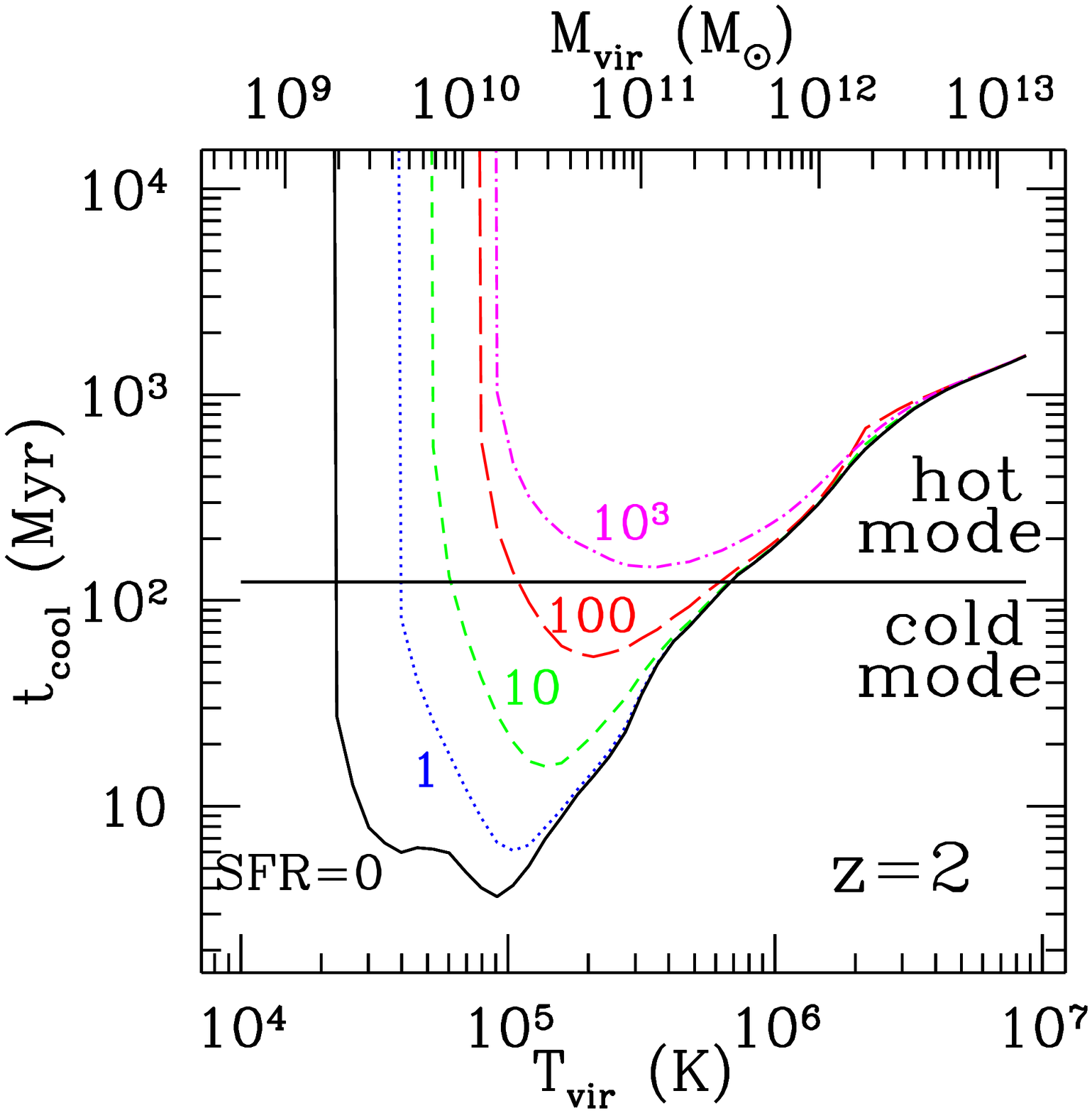,height=55mm}
  \psfig{file=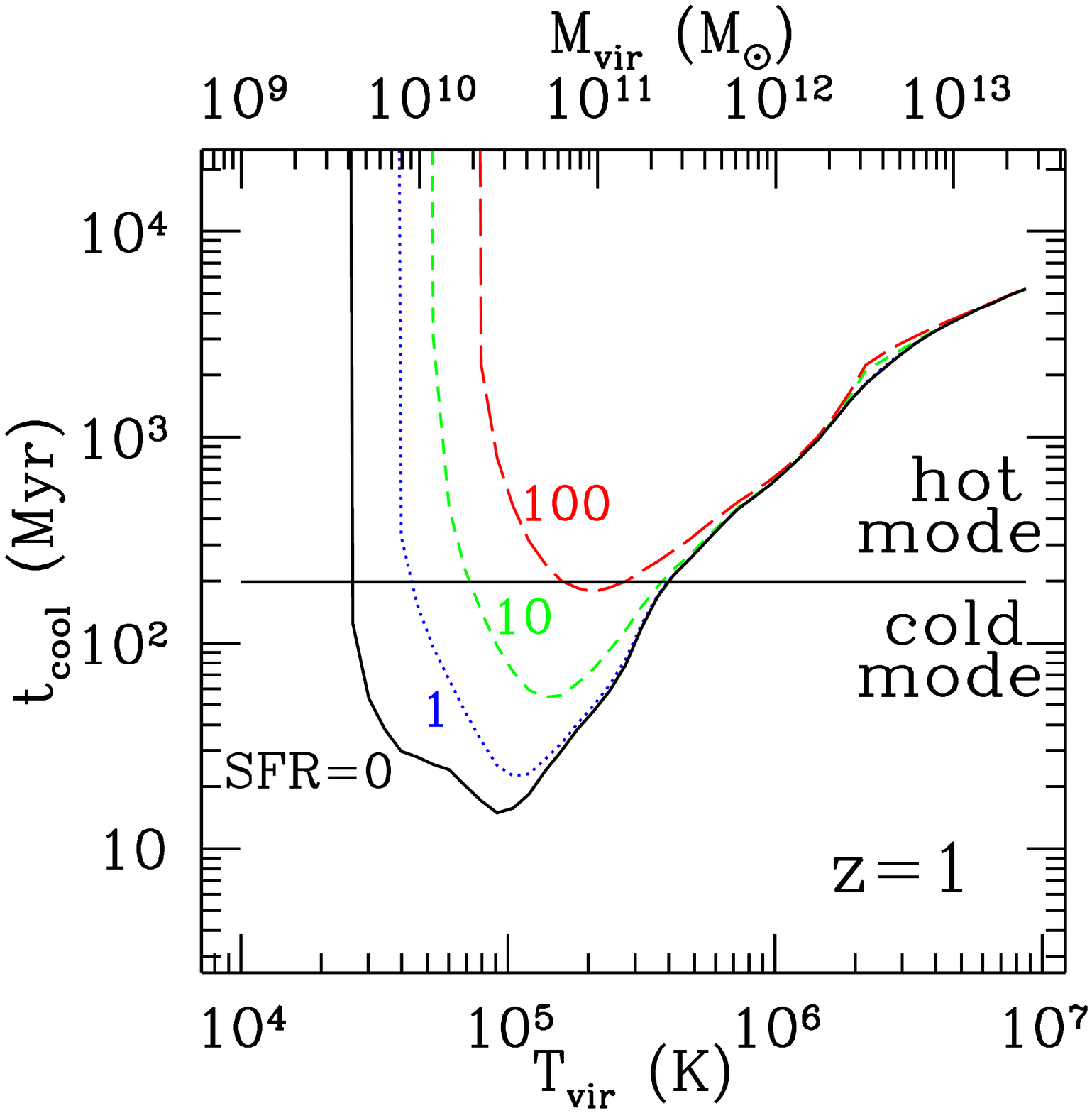,height=55mm}
  \psfig{file=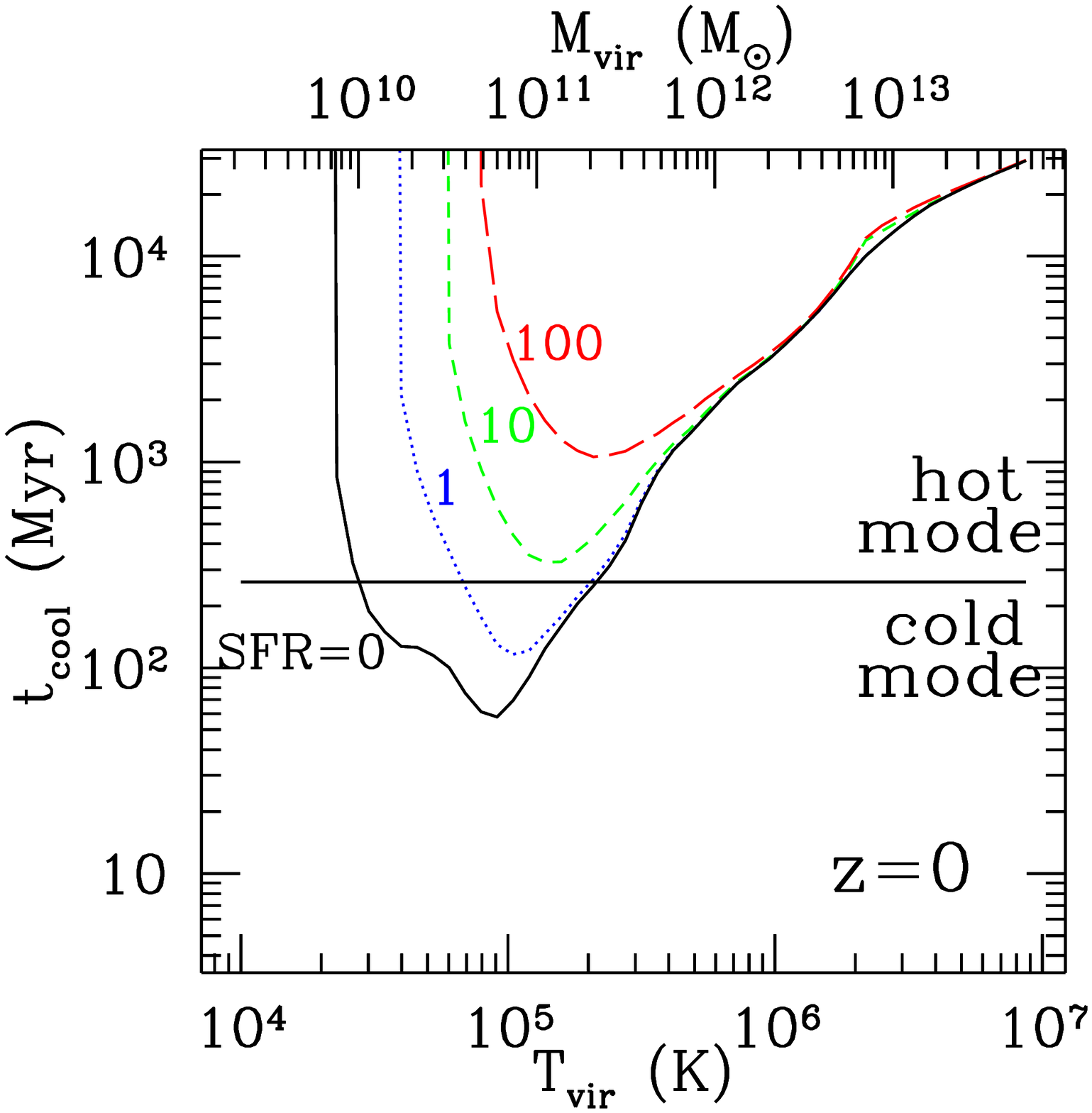,height=55mm}
}}
\caption{Cooling time as a function of gas temperature (or, equivalently, $M_{vir}$) 
in the presence of the cosmic UV background only (SFR$=0$) and for different values of the SFR (in units of  $\mathrm{M}_{\odot}$ yr$^{-1}$) 
and redshifts (see labels).  
The gas density is calculated (at $0.1R_{vir}$) using the halo virial relations taking into account
the redshift evolution and the mass dependence of the concentration parameter (see text for details). The gas metallicity is fixed at $Z=0.03\mathrm{Z_{\odot}}$. 
The horizontal lines give the value of the compression time, $t_{comp}$, defining the transition between ``cold'' and ``hot'' accretion mode
at the corresponding redshift.  
}
\end{figure}

\section{Implication for galaxy formation}

 The cooling time of a patch of gas with density $n_{H}$ and temperature $T$ is defined by:
\begin{equation}
t_{cool}=\frac{3}{2}\frac{n_{\mathrm{H}}k_{\mathrm{B}}T}{\Lambda_{net}(T,n_{\mathrm{H}},Z,U_{*},z)}\ .
\end{equation}
We compare $t_{cool}$ to the compression time of the gas as it falls within the dark-matter halo ($t_{comp}$, 
roughly equal to few times the gas free fall time
),
following Dekel \& Birnboim 2006 (DB06).
Note that the last equation
is valid in the innermost part of the halo (at about 10\% of the virial radius). 
As extensively discussed by DB06, the relation between these two time-scales around $r=0.1R_{vir}$ seems to be the
origin of the transition between ``cold'' and ``hot'' mode accretion.
The actual value of the compression time depends on the details of the shock (e.g., the shock speed) and the gas
profile. Nevertheless, the simple condition $t_{cool}>t_{comp}$ (as defined above) seems to produce very similar results
to the one obtained from numerical simulations (e.g., Ocvirk et al. 2008), where no assumptions
about the shock properties are made, providing that a low gas metallicity ($Z\simeq 0.03 \mathrm{Z_{\odot}}$) is assumed.
Therefore, in the rest of this Letter, we use $Z=0.03\mathrm{Z_{\odot}}$ and $r=0.1R_{vir}$.  

 In Fig. 2, we show the gas cooling time 
as a function of the virial shock temperature
(or virial mass) for three different redshifts (see labels in panels). In each panel, the black solid line represents the
cooling time in the presence of the HM05 only, while the horizontal lines show the value of $t_{comp}$ at the given redshift.
We have computed the gas overdensity at $r=0.1R_{vir}$ using the halo virial relations, assuming
an isothermal profile for the gas, a halo gas fraction $f_b=0.05$ (consistently with DB06), 
and taking into account the evolution of the haloes concentration with redshift and mass (using the fit of Bullock et al. 2001).  
Similarly to DB06, where local sources are not included, 
the transition between ``hot'' and ``cold'' mode happens at the critical halo mass $M_{crit}\sim3\times10^{11}\mathrm{M_{\odot}}$, 
for $Z=0.03\mathrm{Z_{\odot}}$, independently of redshift. 
The other lines in Fig.2 show the cooling time in the presence of the local ionizing flux
for different values of the SFR in units of  $\mathrm{M}_{\odot}$ yr$^{-1}$ (see labels). 
It is clear that (especially at low redshifts) the modality of gas accretion depends on the presence
of star-formation activity: at a given redshift, $M_{crit}$ is actually determined by the value of the SFR.
Note that, at the lower mass end ($M^{vir}_{11}<0.5$), 
cooling is completely suppressed (and the gas is actually slightly heated above $T_{vir}$).

\begin{figure}
\centerline{
\includegraphics[width=75mm]{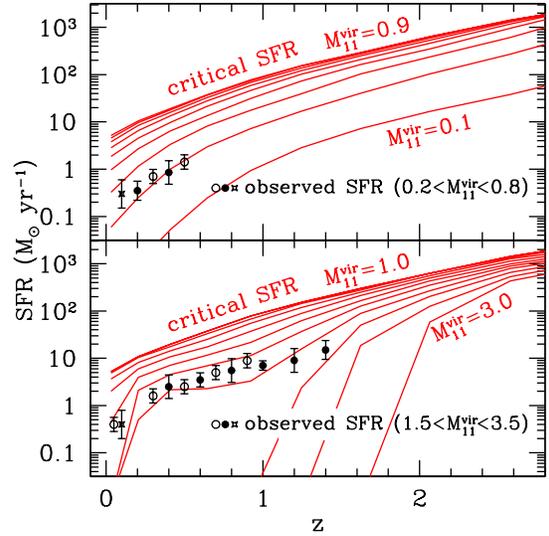}
}
\caption{Evolution of the \emph{critical} SFR (the SFR at which $t_{cool}=t_{comp}$ for $r=0.1R_{vir}$, and ``cold'' mode is stopped) 
for different halo masses (solid red contour lines). The gas density is calculated using the halo virial relations taking into account
the redshift evolution and the mass dependence of the concentration parameter (see text for details). The gas metallicity is fixed 
at $Z=0.03\mathrm{Z_{\odot}}$.
In the upper (lower) panel, contours are equally separated by $0.1$ ($0.2$) in $M^{vir}_{11}$. Note that  the \emph{critical} SFR increases
with halo masses for $M^{vir}_{11}<1$ (upper panel) while the trends is inverted for $M^{vir}_{11}>1$ (lower panel), as expected given the
relation between cooling time and halo masses shown in Fig. 2.
For illustrative purposes, we overlay the observed SFRs of galaxies with
$M_{*}\sim2\times10^9\mathrm{M_{\odot}}$ (upper panel) and $M_{*}\sim10^{10}\mathrm{M_{\odot}}$ (lower panel) obtained by
the SDSS survey (crosses), Zengh et al. 2007 (open circles) and Damen et al. 2009 (circles). 
The observed stellar masses have been converted to an approximate $M_{vir}$ range extrapolating the 
results of Mandelbaum et al.(2006). 
}
\end{figure}

 In Fig. 3, we show the redshift evolution of the \emph{critical} SFR (i.e., the SFR at which $t_{cool}=t_{comp}$ for $r=0.1R_{vir}$
and ``cold'' mode accretion is stopped; solid red lines) for different halo masses. Note that the \emph{critical} SFR increases
with halo masses for $M^{vir}_{11}<1$ (upper panel) while the trends is inverted for $M^{vir}_{11}>1$ (lower panel). This is simply due to
the shape of the cooling function or, equivalently, to the relation between the cooling time and halo mass (see Figure 2).    
For illustrative purposes, we overlay the observed SFRs of galaxies with
total stellar mass of the order of $M_{*}\sim2\times10^9\mathrm{M_{\odot}}$ (upper panel) and $M_{*}\sim10^{10}\mathrm{M_{\odot}}$ (lower panel) from various sources (see caption).
To facilitate the comparison, we converted the stellar masses to an approximate $M_{vir}$ range extrapolating the 
results of Mandelbaum et al. (2006). Note that large uncertainties are associated with this conversion, especially for
the low mass bin. Nevertheless, it appears remarkable that the \emph{critical} SFR (that basically depends only on the 
cooling physics) is close to the observed value. Also, the slope of the redshift evolution
appears very similar to the observed relation.
Note that the \emph{critical} SFR represents the \emph{instantaneous} value of the SFR that is needed in order to stop
``cold'' mode accretion, while the observed values represent the \emph{average} SFRs of the galaxies (that may also be at
different stages of evolution). In this respect, the \emph{critical} SFR should be regarded as a sort of maximum 
and for this reason we might expect the observed values to be below it, as in Fig. 3.

\section{Discussion and conclusions}\label{discussion}

 We have argued that the local photoionizing flux 
produced by star-forming galaxies could play
a fundamental role for the cooling and accretion of the halo gas. 
In particular, we have found that the soft X-rays and EUV photons produced
by star formation activity, are able to remove from the gas important coolants
(e.g., O$^{4+}$ and Fe$^{8+}$) via direct photoionization. The reduced cooling rate
(and, partially, the increased photoheating) result in a much longer cooling time for the
accreting gas. As a consequence, ``cold'' mode accretion may be stopped for haloes well below
the critical halo mass for shock-heating, $M_{crit}\sim3\times10^{11}\mathrm{M_{\odot}}$, 
inferred neglecting local photoionization sources. 

Specifically, we have found  
that (for a given halo mass and redshift) there is a \emph{critical} SFR 
corresponding to the transition between ``cold'' and ``hot'' mode accretion. The evolution
of the critical SFR with redshift, for a given halo mass, is of the form $(1+z)^{\alpha}$ with
$\alpha\sim3$, resembling the corresponding steep evolution of the observed SFR for $z<1$ (e.g.,
Damen et al. 2009). This intriguing result suggests that, at least for this redshift range and for haloes
with $10^{9.5}<(M_{vir}/\mathrm{M_{\odot}})<10^{11.5}$, 
photoionization by local sources may represent an important mechanism 
to regulate gas accretion and star formation 
helping to relieve the need of additional, strong feedback processes (e.g., galactic winds or mechanical SN feedback). 
This may help to solve a long-standing
problem of numerical models of galaxy formation: the effective quenching of star formation in intermediate and low-mass 
haloes. 

How sensitive are these results to variation in metallicity and densities assumed here? Following DB06, we used a simple
model based on the halo virial relations and the evolution of the concentration parameter to derive the
densities that enter into the definition of the cooling time. This approach gives very similar transition masses
between ``cold'' and ``hot'' mode accretion with respect to hydro-dynamical simulations (e.g., Ocvirk et al. 2008).
These models also find that ``cold'' mode accretion is typically associated with low metallicity gas, as 
we have assumed. Nevertheless, single objects may have different densities and metallicities with respect to the ``average'' values
used here. In order to asses what is the effect of an enhanced or decreased density on our results, we
have rescaled the densities given by the halo virial relations by a factor $f_{e}$. We found that the
\emph{critical} SFR scales roughly like $f_{e}^2$ (for $M^{vir}_{11}\sim 1$ and $0.5<f_{e}<5$), similarly
to what expected in the CIE case. In the case of a more metal
enriched gas the \emph{critical} SFR also increases, up to a factor of two for $Z=0.1\mathrm{Z_{\odot}}$.
These effects may introduce a scatter into the relation between the \emph{critical} SFR and halo masses
(changing the normalization of the curves presented in Figure 3), however they do not change its steep evolution
with redshift, that is in good agreement with observations.

What is the effect of additional sources of photoionization and heating? 
 AGN activity is sometimes connected with star-forming galaxies,
boosting by few order of magnitudes the soft-X ray emission (see e.g., Alexander et al. 2005, Georgakakis et al. 2007).
In this respect, the results presented in this Letter should be regarded as the minimum effect due to star-forming galaxies
when the AGN are not present. Moreover, AGN may be also able to strongly reduce cooling in the ``hot'' mode. 
For a tipical luminosity of $L_{\mathrm{softX}}=10^{43}$ erg s$^{-1}$ the corresponding ionization parameter is
high enough to significantly affect the cooling rate also for $T>10^6$ K (see lines with $U_{*}>1$ in Fig. 1). 
This is consistent, e.g., with the findings of Oh (2004), that discussed this effect as a potential mechanism to 
suppress Fe emission lines from cooling-flow clusters.

 Star formation should be followed by the release
of large amount of energy via SN explosions or stellar winds (Dekel \& Silk 1986) 
that we have not included (as a heating source) in our calculations.
However, the effective amount of energy released into the halo gas and the physical
details of the process are affected by large uncertainties connected, e.g., with the mass loading factor
of the galactic winds. 
We stress that the effect discussed in this Letter is substantially different from SN and winds feedback,
even if they are all connected with star-formation activity: SN feedback heats the
gas while photoionization from local sources \emph{reduces} the cooling rate removing coolant agents 
(and also adds energy through photo-heating). These two effects may work together, reducing the critical SFR
or e.g., helping the re-heated gas from SN feedback to remain hot preventing it to cool and to 
fall back onto the galaxy.

Despite the limitations of the simple analytical arguments used here,
photoionization by local sources certainly represents a fundamental ingredient that should be added into
current galaxy-formation models. In a companion paper, we will show how this effect
changes the galaxy properties with the help of cosmological, hydro-dynamical 
simulations. For the time being, the simple estimates presented in this Letter may represent 
a further step towards a better
understanding of the processes that regulate galaxy-formation and produces the galaxies
as we see them today.

\section*{Acknowledgments}

I thank the anonymous referee for useful comments that improved the Letter.
I am grateful to Martin Haehnelt, Regina Jorgenson, Simon Lilly, Piero Madau, Peng Oh and Debora Sijacki for
useful discussions and for their comments on an earlier version of this manuscript.

\label{lastpage}


\begin{thebibliography}{}
\bibitem[Alexander et al.(2005)]{2005Natur.434..738A} Alexander, D.~M. et al. \ 2005, Nature, 434, 738 
\bibitem[Bergvall et al.(2006)]{2006A&A...448..513B} Bergvall, et al.\ 2006, A\&A, 448, 513
\bibitem[Bullock et al.(2001)]{2001MNRAS.321..559B} Bullock, J.~S., et al., 2001, MNRAS, 321, 559 
\bibitem[Cervi{\~n}o et al.(2002)]{2002A&A...392...19C} Cervi{\~n}o, M., Mas-Hesse, J.~M., \& Kunth, D.\ 2002, A\&A, 392, 19  
\bibitem[]{} Crain, R.~A., et al.\ 2009, MNRAS, 1262 
\bibitem[Damen et al.(2009)]{2009arXiv0908.1377D} Damen, M., et al., 2009, arXiv:0908.1377 
\bibitem[]{} Dekel, A., \& Silk, J.\ 1986, ApJ, 303, 39  
\bibitem[Dekel \& Birnboim(2006)]{2006MNRAS.368....2D} Dekel, A., \& Birnboim, Y.\ 2006, MNRAS, 368, 2, (DB06)  
\bibitem[Efstathiou(1992)]{1992MNRAS.256P..43E} Efstathiou, G.\ 1992, MNRAS, 256, 43P  
\bibitem[Ferland \& Mushotzky(1982)]{1982ApJ...262..564F} Ferland, G.~J., \& Mushotzky, R.~F.\ 1982, ApJ, 262, 564 
\bibitem[Ferland et al.(1998)]{1998PASP..110..761F} Ferland, G.~J., et al.\ 1998, PASP, 110, 761 
\bibitem[Georgakakis et al.(2007)]{2007MNRAS.377..203G} Georgakakis, A., et al. \ 2007, MNRAS, 377, 203 
\bibitem[Grimm et al.(2003)]{2003MNRAS.339..793G} Grimm, H.-J. et al., 2003, MNRAS, 339, 793 
\bibitem[Grimes et al.(2009)]{2009ApJS..181..272G} Grimes, J.~P., et al.\ 2009, ApJS, 181, 272 
\bibitem[Heckman et al.(1995)]{1995ApJ...448...98H} Heckman, T.~M., et al., 1995, ApJ, 448, 98 
\bibitem[Mandelbaum et al.(2006)]{2006MNRAS.368..715M} Mandelbaum, R., et al.\ 2006, MNRAS, 368, 715 
\bibitem[Mas-Hesse et al.(2008)]{2008A&A...483...71M} Mas-Hesse, J.~M., Ot{\'{\i}}-Floranes, H., \& Cervi{\~n}o, M.\ 2008, A\&A, 483, 71  
\bibitem[Mo \& White(2002)]{2002MNRAS.336..112M} Mo, H.~J., \& White, S.~D.~M.\ 2002, MNRAS, 336, 112 
\bibitem[Ocvirk et al.(2008)]{2008MNRAS.390.1326O} Ocvirk, P., Pichon, C., \& Teyssier, R.\ 2008, MNRAS, 390, 1326 
\bibitem[Oh(2004)]{2004MNRAS.353..468O} Oh, S.~P.\ 2004, MNRAS, 353, 468 
\bibitem[Persic et al.(2004)]{2004A&A...419..849P} Persic, M., et al., 2004, A\&A, 419, 849 
\bibitem[Rees \& Ostriker(1977)]{1977MNRAS.179..541R} Rees, M.~J., \& Ostriker, J.~P.\ 1977, MNRAS, 179, 541 
\bibitem[Rosa Gonz{\'a}lez et al.(2009)]{2009MNRAS.399.487} Rosa Gonz{\'a}lez, et al.\ 2009, MNRAS, 399, 487  
\bibitem[Shapley et al.(2006)]{2006ApJ...651..688S} Shapley, A.~E., et al.\ 2006, ApJ, 651, 688 
\bibitem[Silk(1977)]{1977ApJ...211..638S} Silk, J.\ 1977, ApJ, 211, 638  
\bibitem[Steidel et al.(2001)]{2001ApJ...546..665S} Steidel, C.~C., Pettini, M., \& Adelberger, K.~L.\ 2001, ApJ, 546, 665 
\bibitem[Strickland et al.(2004)]{2004ApJS..151..193S} Strickland, D.~K., et al., 2004, ApJS, 151, 193 
\bibitem[Wiersma et al.(2009)]{2009MNRAS.393...99W} Wiersma, R.~P.~C., Schaye, J., \& Smith, B.~D.\ 2009, MNRAS, 393, 99 
\bibitem[Zheng et al.(2007)]{2007ApJ...661L..41Z} Zheng, X.~Z., et al., 2007, ApJL, 661, L41 

 
\end{thebibliography}
\end{document}